\documentclass[10pt,letterpaper,twocolumn]{article} 

\usepackage{ol2}
\usepackage[draft]{hyperref}
\usepackage{amsmath}

\begin{document}

\twocolumn[ 

\title{$N$ -resonances in a buffered micrometric Rb cell: splitting in a strong
magnetic field}

\author{Armen Sargsyan$^1$, Rafayel Mirzoyan$^{1,2}$, Aram Papoyan$^{1*}$, and David  Sarkisyan$^1$}

\address{
$^1$Institute for Physical Research, NAS of Armenia, \\  Ashtarak-2, 0203, Armenia\\
$^2$Laboratoire Interdisciplinaire Carnot de Bourgogne, UMR CNRS 6303, Universite$'$  de Bourgogne,\\ 21078 Dijon Cedex, France \\
$^*$Corresponding author: papoyan@ipr.sci.am
}

\begin{abstract}$N$ -resonances excited in rubidium atoms confined in micrometric-thin cells
 with variable thickness from 1 $\mu$m to 2 mm are studied experimentally for the cases of
  a pure Rb atomic vapor and of a vapor with neon buffer gas. Good contrast and narrow
  linewidth were obtained for thicknesses as low as 30 $\mu$m. The higher amplitude
   and sharper profile of $N$-resonances in the case of a buffered cell was exploited
    to study the splitting of the $^{85}$Rb D$_1$ $N$-resonance in a magnetic field
     of up to 2200 G. The results are fully consistent with the theory. The mechanism
      responsible for forming $N$-resonances is discussed. Possible applications are
       addressed.\end{abstract}


 ] 

\begin{figure}[tbh]
\centerline{\includegraphics[width=5.1cm]{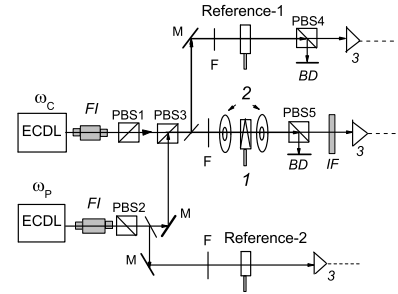}} 
\caption{Sketch of the experimental setup. ECDL- diode lasers; FI- Faraday
isolator; 1- MTC in the oven; PBS- polarizing beam splitters; 2- permanent
ring magnets; 3- photodetectors; IF- interference filter with 10~nm
transmission bandwidth at 795~nm; F- neutral density filters; BD- ~beam dump
to block $\protect\nu _{c}$. PBS5 is used to single out $\protect\nu _{p}$
for detection.}
\label{fig1}
\end{figure}

\begin{figure}[tbh]
\centerline{\includegraphics[width=5.1cm]{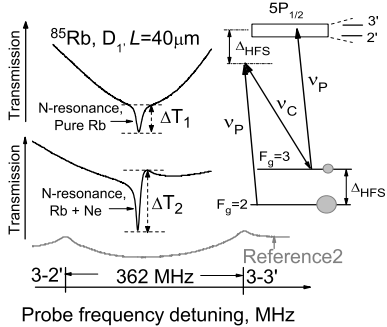}} 
\caption{Transmission spectra of the probe radiation through the MTC with $L$
= 40~$\protect\mu $m. Spectra containing an N-resonance are presented for
two cases: pure Rb vapor (upper curve), which gave a linewidth of around 10
MHz, and Rb with 150 Torr Ne (lower curve), which gave a linewidth of around
8 MHz. Spectra were obtained under nearly identical conditions. For
convenience, the spectra are shifted in the vertical direction. The lower
grey curve is spectrum of Reference-2. Inset: relevant energy levels of $%
^{85}$Rb involved in $N$-resonance formation.}
\label{fig2}
\end{figure}

\noindent The formation of narrow optical resonances via coherent processes
continues to be of high interest because its fascinating properties have
potentially significant applications in quantum information, metrology,
magnetometry, and other fields \cite{1}. Though narrow resonances are
predominantly formed by electromagnetically induced transparency (EIT) and
the related phenomenon of coherent population trapping (CPT) in a $\Lambda $%
-system, other schemes are exploited as well: EIT in a \textquotedblleft
ladder\textquotedblright\ system, electromagnetically induced absorption
(EIA), etc. It was shown in \cite{2} that a so-called $N$-resonance that
forms in a $\Lambda $-system of the Rb D$_{1,2}$ lines and manifests itself
by an increase of the absorption (as it is for the EIA-resonance), can be
competitive with the aforementioned resonances. The main advantage of this
process is that it makes it technically easier to form a high-contrast,
sub-natural resonance. In \cite{3} it was demonstrated that better contrast
could be attained for a D$_{2}$ $N$-resonance, while the linewidth was
narrower for D$_{1}$ $N$-resonance. The possibility of light-shift
cancellation, which might be important for atomic clock applications, was
shown in \cite{4}. The profile asymmetry of $N$-resonances was studied in 
\cite{5}. The parameters of $N$-resonances can be improved by using laser
radiation from three sources \cite{6}. For applications in spectroscopy,
metrology etc., it is important to reduce the dimensions of the atomic vapor
cell in which optical resonances are formed while maintaining good resonance
parameters \cite{7,8,9,10}. In this letter we report on $N$-resonances
observed with wedged micrometric-thin cells (MTC) filled with Rb vapor
together with 150 Torr of neon gas in magnetic fields of up to 2.2 kG. The
thicknesses L of the MTCs ranged from 1 $\mu $m to 50 $\mu $m and from 40 $%
\mu $m to 2 mm. The design of the MTCs used was similar to that presented in 
\cite{10}. The $N$-resonance was formed in a $\Lambda $-system by two lasers
beams with $\lambda \approx $ 795 nm and 1 MHz linewidth: the probe beam,
whose frequency could be tuned, and the coupling beam, whose frequency was
fixed. The diagram presented in the inset of Fig.~\ref{fig2} shows the $%
\Lambda $-system for the $^{85}$Rb atom, where F$_{g}$ = 2,3 are the ground
levels, and the combined upper level 5P$_{1/2}$ consists of hyperfine levels
F$_{e}$=2,3. The probe laser frequency $\nu _{P}$ was resonant with the 5S$%
_{1/2}$, F$_{g}$=3 $\longrightarrow $ 5P$_{1/2}$ transition, and the
coupling laser frequency was shifted by the value of the ground state
hyperfine splitting ($\Delta _{HFS}):\nu _{C}=\nu _{P}-\Delta _{HFS}$. The
experimental arrangement is sketched in Fig.~\ref{fig1}. 

The beams of two single-frequency extended cavity diode lasers (ECDL) were
carefully superimposed and directed by PBS3 onto the MTC. The coupling and
probe beams were linearly polarized in orthogonal planes. The small
thickness of MTC makes it possible to use a permanent ring magnet (PRM) in
order to apply a strong magnetic field and still obtain a homogeneous field
over the thickness of the cell: in the MTC, the variation of the $B$-field
inside the atomic vapor column is negligible compared to value of the
applied magnetic field. The magnetic field was measured by a calibrated Hall
gauge. To control the magnetic field value, one of the magnets was mounted
on a micrometric translation stage that allowed longitudinal
displacement.Portions of the coupling and probe beams were diverted to an
auxiliary 40~$\mu $m-long Rb cell filled with Ne gas to obtain an $N$%
-resonance spectrum at $B$ = 0. 
This spectrum served as frequency Reference-1. Furthermore, another portion
of the probe beam was diverted to a Rb nano-cell with $L$~=~$\lambda $ to
obtain a $B$~=~0 transmission spectrum, which served as frequency Reference
2 \cite{11}. The optical radiation signals recorded by photodiodes (3) were
amplified and recorded by a four-channel digital storage oscilloscope.

Although the best $N$-resonance contrast and linewidth can be achieved for
cells with thicknesses around 1~cm \cite{2,3,4,5,6}, using an MTC with a
thickness as small as 30 to 40~$\mu $m still allowed us to obtain good
resonance parameters. The experimentally recorded $N$-resonance spectra are
presented in Fig.~\ref{fig2}. The MTC side arm, whose temperature determines
the density of Rb atoms, was maintained at $\sim 110^{\circ }$C (Rb atomic
vapor density 10$^{13}$~cm$^{-3}$). In the case of the pure Rb vapor (the
upper curve) the change of the probe transmission over the $N$-resonance was 
$\Delta T_{Rb}\approx 7\%$, and its lineshape was symmetric. For the
buffered cell (the lower curve) the change was larger ($\Delta
T_{Rb+Ne}\approx 12\%$), and the $N$-resonance shape was asymmetric,
consistent with the results reported in \cite{5}. However, the sharp profile
of the transmission signal in this case makes it convenient for studying the
splitting of $N$-resonances in a magnetic field. 
\begin{figure}[tbh]
\centerline{\includegraphics[width=5.1cm]{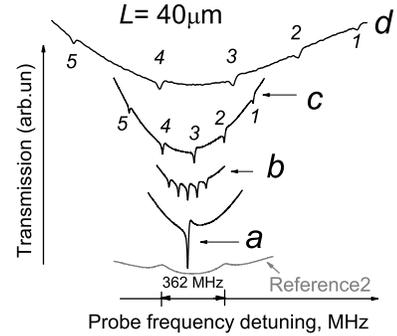}} 
\caption{Splitting of the $N$-resonance in a moderate $B$-field. a: spectrum
of Reference-1 for $B$~=~0; \emph{b - d}: $N$-resonance spectra for $B$ =
59~G (\emph{b}), 190~G (\emph{c}), and 460~G (\emph{d}). The labels \emph{1-5%
} denote corresponding transitions shown in Fig.~\protect\ref{fig6}. The
lower grey curve shows spectrum of Reference-2. }
\label{fig3}
\end{figure}
The $N$-resonance was split into 5 components in a magnetic field. This
splitting is shown in Fig.~\ref{fig3} for the $B$-field range from 59~G to
460~G. The coupling and probe beam powers were 4~mW and 1~mW, with a beam
diameter of 1.5~mm.

\begin{figure}[htb]
\centerline{\includegraphics[width=5.5cm, height=4.5cm]{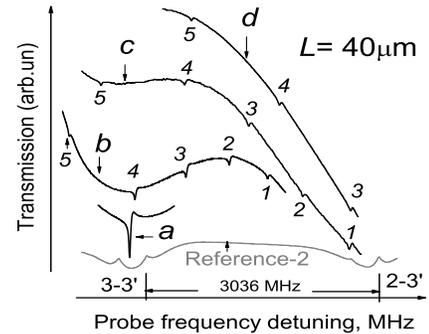}} 
\caption{Splitting of the $N$-resonance in a strong $B$-field. a: spectrum
of Reference-1 for $B$ = 0; \emph{b - d}: $N$-resonance spectra for
B~=~808~G (\emph{b}), 1238~G~(\emph{c}), and 1836~G~(\emph{d}). The labels 
\emph{1-5} denote corresponding transitions shown in Fig.~\protect\ref{fig6}%
. The lower grey curve shows spectrum of Reference-2. }
\label{fig4}
\end{figure}

Spectra showing splitting for stronger $B$-field values (808 - 1836~G) are
presented in Fig.~\ref{fig4}. Although the amplitudes of the $N$-resonance
components tended to decrease with $B$, they were nevertheless easily
observable up to $B$ = 2200~G. As can be seen from Fig.~\ref{fig3} and Fig.~%
\ref{fig4}, the narrow linewidth of the $N$-resonance makes it possible to
achieve a spectral resolution that is higher by a factor of 5 as compared
with the results obtained by the $\lambda $-Zeeman technique \cite{9}.

\begin{figure}[htb]
\centerline{\includegraphics[width=5.1cm]{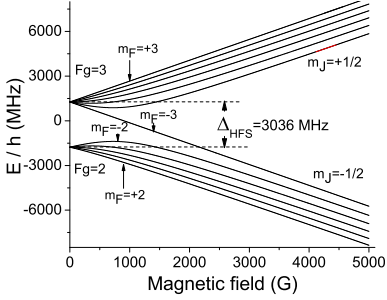}} 
\caption{Splitting of $^{85}$Rb 5S$_{1/2}$ ground level hyperfine structure
in an external magnetic field.}
\label{fig5}
\end{figure}

Figure~\ref{fig5} presents the dependence of the frequency of the magnetic
sublevels of the $^{85}$Rb F$_{g}$=2,3 ground hyperfine states on the
magnetic field, as calculated by a well-known model (see, for example, \cite%
{11}). The system is described in the basis of F, m$_{F}$ in the low-field
(Zeeman) regime, and in the basis of m$_{J}$, m$_{I}$ in the strong-field
(hyperfine Paschen-Back [HPB]) regime, when B~$\gg $~700~G \cite{9}. $N$%
-resonance components are observed whenever the 2-photon resonance
conditions are satisfied: $\nu _{P}-\nu _{C}$~$=$~$[E$(F=3, m$_{F}$)~$-$~$E$%
(F=2, m$_{F})]/h$.

\begin{figure}[htb]
\centerline{\includegraphics[width=6.1cm, height=4.7cm]{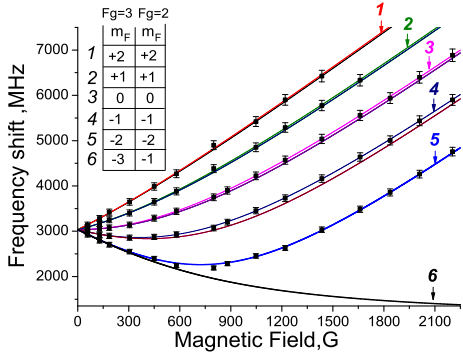}} 
\caption{Frequency shifts of the $N$-resonance components in a $B$-field. $%
Solid$ $lines$: theory; $symbols$: experiment. The inaccuracy does not
exceed 2$\%$. The initial and final Zeeman sublevels of F$_g$=2,3 are
indicated in the Table for components \emph{1-5}. }
\label{fig6}
\end{figure}

Figure~\ref{fig6} shows the magnetic field dependence of the frequency shift
for the five observed $N$-resonance components. Other $N$-resonance
components expected from the calculations and presented as thin solid lines
in Fig.~\ref{fig6} are not observable in the experiment because at low
fields ($B<$ 500 G) their separation from components \emph{1-5} is
unresolvably small (several MHz), while at high fields the atomic transition
probabilities (line strengths) are reduced (either for the coupling or for
the probe radiation) down to undetectable values (see component 6,
undetectable at $B>$ 500 G). Note that the slope of components \emph{1-5}
contains contributions from the $B$-field shifts of the corresponding
ground-state Zeeman sublevels F$_{g}$=2,3 (see Fig.~\ref{fig5}). Thus, $N$%
-resonance component \emph{1} has the largest slope of 2.68~MHz/G in the
region around 2~kG, that is, $\left\vert -1.34\right\vert $~MHz/G from F$_{g}
$=2, m$_{F}$=+2, and $\left\vert +1.34\right\vert $~MHz/G from F$_{g}$=3, m$%
_{F}$=+2. For $B$ $>$ 1.5~kG an HPB regime becomes predominant. As a
consequence, the slopes of components \emph{1-5} tend to equalize at the
same final value of $\sim $ 2.8~MHz/G .

The studies presented above indicate that F$_{g}$=2 is the initial, and F$%
_{g}$=3 is the final level for the $N$-resonance. Based on these results, we
believe the following physical mechanism is responsible for the origin of
the $N$-resonance.The probe radiation causes strong optical pumping, which
transfers a large number of Rb atoms from the F$_{g}$=3 state to the F$_{g}$%
=2 state. The presence of buffer gas enhances efficiency of optical pumping,
yet weakly influences the spectral broadening (at the given pressure). In
consequence, an $N_{2}~>$~$N_{3}$ population condition is assured (where N$%
_{2}$ and N$_{3}$ are the populations of the F$_{g}$=2 and F$_{g}$=3 levels,
respectively). This condition is schematically indicated by the large and
small circles in the inset of Fig.~\ref{fig2}. When the condition $\nu
_{p}~-~\nu _{c}~=~\Delta _{HFS}$ is satisfied, a strong two-photon
absorption of the probe radiation F$_{g}$=2 $\longrightarrow $ F$_{g}$=3
occurs at the frequency $\nu _{p}$~$=$~$\nu _{C}~+\Delta _{HFS}$ , which
results in the formation of an $N$-resonance.

Note that the 30 $\mu$m-thick Rb cell could be used to map strongly
inhomogeneous magnetic fields with high spatial resolution. In particular,
for a $B$-field gradient of around 100 G/mm, the displacement of the MTC by
30~$\mu$m results in a frequency shift of the $N$-resonance component
labeled \emph{1} of around 8~MHz, which is easy to detect because of its
sharp profile.

We thank F. Gahbauer for useful discussions.

\pagebreak

\end{document}